\documentclass[sigconf]{acmart} 
\AtBeginDocument{%
  \providecommand\BibTeX{{%
    \normalfont B\kern-0.5em{\scshape i\kern-0.25em b}\kern-0.8em\TeX}}}

\renewcommand\footnotetextcopyrightpermission[1]{} % removes footnote with conference information in first column

\usepackage{xspace}
\usepackage{subfigure}
\usepackage{balance}

\usepackage{listings}
\usepackage{color}
\usepackage{wrapfig}
\definecolor{pyred}{rgb}{0.6,0,0} % for strings
\definecolor{pygreen}{rgb}{0.25,0.5,0.35} % comments
\definecolor{pypurple}{rgb}{0.5,0,0.35} % keywords
\definecolor{pydocblue}{rgb}{0.25,0.35,0.75} % javadoc
 
\newcommand{\fakepar}[1]{\vspace{.5mm}\noindent\textbf{#1.}}
\newcommand\figref[1]{Fig.\,\ref{#1}}
\newcommand\secref[1]{Sec.\,\ref{#1}}

\newcommand\appref[1]{Appendix\,\ref{#1}}
\newcommand{\cf}{Crazyflie\xspace}

\newcommand{\code}[1]{\texttt{\bf#1}}

\begin{document}

\title{Shaping and Being Shaped by Drones: \\ Supporting Perception-Action Loops}

\author{Mousa Sondoqah$^*$, Fehmi Ben Abdesslem$^+$, Kristina Popova$^\dagger$, 
Moira McGregor$^\ddagger$, Joseph La Delfa$^{\dagger\ddagger}$, Rachael Garrett$^\dagger$,
Airi Lampinen $^\ddagger$, Luca Mottola$^{*+}$, and Kristina Höök$^+\dagger$\\
$^*$Politecnico di Milano (Italy), $^+$RI.SE (Sweden),\\ $^\dagger$KTH
Royal Institute of Technology (Sweden), $^\ddagger$ Stockholm
University (Sweden)}
%%\email{trovato@corporation.com}
%%\orcid{1234-5678-9012}
%%\author{G.K.M. Tobin}
%%\authornotemark[1]
%%\email{webmaster@marysville-ohio.com}
%%\affiliation{%
%%  \institution{Institute for Clarity in Documentation}
%%  \streetaddress{P.O. Box 1212}
%%  \city{Dublin}
%%  \state{Ohio}
%%  \country{USA}
%%  \postcode{43017-6221}
%%}

%%\renewcommand{\shortauthors}{Trovato and Tobin, et al.}

\begin{abstract}

We report on a three-day challenge during which five teams each programmed a nanodrone to be piloted through an obstacle course using bodily movement, in a 3D transposition of the '80s video\-game Pacman. Using a bricolage approach to analyse interviews, field notes, video recordings, and inspection of each team's code revealed how participants were shaping and, in turn, became shaped in bodily ways by the drones' limitations. We observed how teams adapted to compete by: \emph{1)} shifting from aiming for seamless human-drone interaction, to seeing drones as fragile, wilful, and prone to crashes; \emph{2)} engaging with intimate, bodily interactions to more precisely understand, probe, and delimit each drone's capabilities; \emph{3)} adopting different strategies, emphasising either training the drone or training the pilot. We contribute with an empirical, somaesthetically focused account of current challenges in HDI and call for programming environments that support action-feedback loops for design and programming purposes. %Drawing upon an analogy of the evolution of personal computing, we argue for direct-manipulation as a path to drone technology programming and creating for a continuous development of drone behaviors based on the insights we gained during the challenge.

\end{abstract}

\maketitle
\pagestyle{empty} % removes running headers

%%\begin{CCSXML}
%%<ccs2012>
%% <concept>
%%  <concept_id>10010520.10010553.10010562</concept_id>
%%  <concept_desc>Computer systems organization~Embedded systems</concept_desc>
%%  <concept_significance>500</concept_significance>
%% </concept>
%% <concept>
%%  <concept_id>10010520.10010575.10010755</concept_id>
%%  <concept_desc>Computer systems organization~Redundancy</concept_desc>
%%  <concept_significance>300</concept_significance>
%% </concept>
%% <concept>
%%  <concept_id>10010520.10010553.10010554</concept_id>
%%  <concept_desc>Computer systems organization~Robotics</concept_desc>
%%  <concept_significance>100</concept_significance>
%% </concept>
%% <concept>
%%  <concept_id>10003033.10003083.10003095</concept_id>
%%  <concept_desc>Networks~Network reliability</concept_desc>
%%  <concept_significance>100</concept_significance>
%% </concept>
%%</ccs2012>
%%\end{CCSXML}

%%\ccsdesc[500]{Computer systems organization~Embedded systems}
%%\ccsdesc[300]{Computer systems organization~Redundancy}
%%\ccsdesc{Computer systems organization~Robotics}
%%\ccsdesc[100]{Networks~Network reliability}

%%\keywords{datasets, neural networks, gaze detection, text tagging}


\section{Introduction}

Aerial nano-drones are a fascinating and evocative type of robot. They have a material body~\cite{despret2013responding} that moves, they make noise, and they can tempt us to interact with them, as if they were more communicative than they are. Design processes focused on human–drone interaction (HDI) feature encounters between the designers, with their designerly aims and fleshy bodies, and drones with unpredictable technological affordances and fragile plastic bodies~\cite{popova2022vulnerability}. As these encounters unfold, designers shape not only the technology but also themselves, especially their movements~\cite{eriksson2019dancing,la2020designing}. 

We study what happens when people with little to no prior experience of programming and piloting drones, or interaction design, are tasked with shaping HDI; we report on a project where we organized, documented, and analyzed a three-day drone challenge. Participating teams programmed nano-drones to be piloted through an obstacle course using bodily gestures, in a 3D transposition of the '80s Pacman videogame. Participants were given a limited time for exploring interactions with the drone, and they were tasked to work toward the specific objective of piloting the drone through the obstacle course quickly and precisely. This setting differs drastically from prior open-ended designerly explorations~\cite{eriksson2019dancing, la2020drone, popova2022vulnerability,gamboa2022living}.

In contrast to prior work that concerns human-drone observation \cite{vanwaveren2023increasing, kaduk2023effects, randall2023picture} or human-drone gestures %\footnote{Here we use the word \textit{gesture} as a classified set of movements interpreted as a command, later in the paper we use the word in a different sense to describe the actions of the team members.} 
\cite{hosseini2023towards, wilsonsmall2023teacher, heesoon2018aeroquake}, we take as our starting point the inseparable coupling between perception and action when controlling a drone with bodily inputs \cite{huppert2021guide,10.1145/3569009.3572740,tsykunov2019swarmcloak}. Our analysis is focused on \textit{the mutual shaping of pilots and drones}, that is, how participants were shaping and, in turn, became shaped in bodily ways by the drones' limitations. We approach this theme as a multidisciplinary team that combines competence in interaction design, human-computer interaction, embedded systems, and mobile robotics. Our bricolage analysis draws upon interviews, field notes featuring both observations and organizers' own experiences, video recordings, and inspection of the teams' programs.
Our findings highlight three issues.
\begin{enumerate}
\item We depict how participants' understanding of drones transformed from inspirational imagery of seamless HDI to lived experience of drones as fragile, wilful, and prone to crashes.
In the process of understanding the sources of these issues, participants faced the \emph{visibility problem}~\cite{lee2016introduction}, common in embedded programming: the internal states of the system are not visible from the outside and it is consequently hard to understand why the drone is behaving in unexpected ways. 
%Narratives about drones tend to ignore or undermine the prevalence of problems and crashes, but in reality they take up much of the time that people interact with drones. 
\item We describe how participants came to know their drones better through intimate, bodily interactions to more precisely probe, understand, and determine the capabilities of their drone. 
This process of getting to know the drone unfolded not only through collaboration among team members but, importantly, with the support of the organizing team who provided technical support. 

\item Through discussing teams' differing approaches to building up their drone interactions, we illustrate the difference between \textit{training the drone}, for instance by applying \textit{defensive coding} in the form of programs that operate cautiously in anticipation of possible issues, and \textit{training the pilot}, that is, changing and rehearsing the pilot's own movements to fit with existing drone capabilities. 
\end{enumerate}

We contribute with an empirical, somaesthetically focused account of current challenges in HDI and a discussion on how we might better support somatic engagement in the design and development of drones. %In an analogy to the evolution of personal computing, we argue that forms of direct-manipulation in the programming of drones are needed. As discussed by Mueller~\cite{} \textit{"computing started with machines that executed a program in one go before returning the result to the user. By decreasing the interaction unit to single requests, turn-taking systems such as the command line evolved, which provided users with feedback after every input. Finally, with the introduction of direct-manipulation interfaces, users continuously interacted with a program receiving feedback about every action in real-time"}. 
%We draw on this analogy to call for ways to directly-manipulate drones' sensing and actuation mechanisms in and through movement, as a way of programming their behaviour. 
To understand drones, we cannot focus solely on either watching their behavior or on figuring out how they sense human movement. Rather, what is needed is entering a \textit{perception-action loop}: it is through movement that we come to understand and act in the world. We call for programming environments that elucidate and enable action-feedback loops for design and programming purposes. %For example, embedded programming tools for drones do not only need to allow programmers to adequately change the code or sensor filters of the drones. Drones must also "train" their pilots through providing somatic signs and signals that can be probed and felt in real-time~\cite{la2020drone}. This would allow pilots to learn how they might shape their own movements to better pilot the drone, until they develop a new way of moving together with the drone. 
Supporting action-feedback loops, by making them somatically felt as they unfold over time, would enable designers to better work with drones and other robots. It would allow them to get to know the technologies they are interacting with and, thus, enable them to shape technologies in line with their aims.

\section{Background}
\label{sec:background}

We provide a brief overview of drone piloting from a technical standpoint. We then discuss prior research on HDI, with a somaesthetic focus on movement-based explorations.

%and how the unfolding relationship between people and drones may go past conventional piloting actions.

%\begin{figure}[tb]
%  \centering
%  \subfigure[Drone piloting approaches.]{
%    \label{fig:arch}
%\includegraphics[height=5cm, keepaspectratio]{pics/droneArch}
%  } \hspace{1cm}
%  \subfigure[Handheld drone controller.]{
%      \label{fig:controller}  
%  \includegraphics[height=5cm, keepaspectratio]{pics/rcController.jpeg}
%  }
%   \caption{Drone piloting technology.}
%   \label{fig:mitreo}
%\end{figure}

%\begin{figure}[tb]
%  \centering
%    \includegraphics[height=4.8cm, keepaspectratio]{pics/droneArch}
%   \caption{Drone piloting approaches.}
%   \label{fig:arch}
%   \vspace{-3mm}
%\end{figure}

\subsection {Drone Piloting}

Conventionally, drone piloting may be achieved in three ways:

First, a control station may wirelessly connect to the drone to issue high-level piloting commands, such as "move forward 1 m", or "rotate $90^{\circ}$ right".
The control station runs a set of programs that implements arbitrary application logic, for example, to achieve a given coverage of a geographical area, achieving \emph{fully autonomous} behaviors~\cite{floreano2015science}. 
Most professional drone platforms operate this way.

Alternatively, a piloting device can be \emph{manually} operated by a trained pilot, who has full control of the drone actions.
Manual drone piloting is challenging. 
The set of knobs and handles available on a piloting device is both small and large.
It is small in that modern drones offer a multitude of operating modes and each such mode may require a separate set of knobs and handles.
It is, at the same time, large for a human to control in real time while maintaining eye contact on the drone, as dictated by current regulations.

Halfway between the two extremes lie the many solutions that offer some form of \emph{assisted} drone piloting.
Some form of piloting device is used here as well, like a mobile app on smartphones.
The piloting device offers ways to achieve some degree of manual control, while the drone autonomously performs other piloting actions in support, such as keeping the position hovering if no piloting input is received.
Most consumer and pro-sumer drone platforms adopt this approach.

The inputs received from either the control station or a piloting device are processed by a piece of embedded software, running aboard the drone, called \emph{flight controller}~\cite{bregu2016reactive}. 
Its job is to realize the movements requested by the control station or by the pilot, translating their inputs into operational settings for the motors. 

\subsection {Somaesthetic Human-Drone Interactions}

Tezza and Andujar~\cite{tezza2019state} describe how drones expanded from military operations to a range of civilian applications. The social-cultural implications~\cite{miah2020drones} of this expansion prompted a surge of research into HDI~\cite{flyinginterfaces,cauchard2021toward}, particularly in the design space of social drones~\cite{socialdrones}. Much research into the relational aspects~\cite{hildebrand2021aerial} of drone technology in these settings has focused on so-called "natural" HDI~\cite{naturalhdi}, integrating gestures and movements \cite{droneandme}, and studying emotions in relation to the design of such systems~\cite{lovedrones, emotiondrones, vanwaveren2023increasing}. Drawing upon Communication Studies to advance HRI, Urakami and Seaborn~\cite{10.1145/3570169} have suggested a series of nonverbal codes that address the five human sensory systems  to promote more natural, inviting, and accessible experiences.%It is argued that, with the increasing prevalence of consumer drones, it is important to achieve a "natural" human-drone interaction~\cite{tezza2019state}.

Joining scholars who have challenged any straightforward notion of naturalness~\cite{hook2018designing, suchman1997interactions, dourish2001action, norman2010natural}, we approach HDI with a somaesthetic sensibility. We are driven by the concept of a ‘soma’ or the lived and felt body as it exists, moves, and senses in the world~\cite{shusterman2008body}. %Somaesthetics as a theory provides an ethical stance on the soma in that it draws attention to how technologies and interactions encourage certain movements and practices, while discouraging others. 
We consider a somaesthetic perspective generative for analyzing human–drone interaction as it invites us to address the limited and limiting ways in which we move~\cite{hook2018designing, sheets1999emotion}. When we interact closely with drones, we have to adapt ourselves in how we control them and move around them. This happens when using them for work purposes \cite{khan2019exploratory, silvagni2017multipurpose}, as part of leisure activities \cite{karjalainen2017social, 10.1145/3434074.3447148}, artistic performances \cite{kov, eriksson2019dancing, heesoon2018aeroquake}, or in family settings~\cite{gamboa2022living, 10.1145/3434074.3446363}. Here, we draw on prior work on HDI in artistic performance~\cite{eriksson2019dancing, eriksson2020ethics}, Tai Chi-inspired movement~\cite{la2020drone, la2020designing}, and design processes~\cite{popova2022vulnerability, 10.1145/3569009.3572740}. 

These prior works offer scaffolding for studying movement-based interactions with drones, as they consider both the somatic aspects of human–drone interaction  as well as the social settings where such interaction takes place. For example, Popova et al.~\cite{popova2022vulnerability} explored possibilities for designing relationships with “drones as ‘the other’ – a distinctive and separate entity, which is neither completely controlled, nor fully autonomous”. %The authors were curious to see if movement, dance, or singing could allow them to direct, influence, or interact with drones as if they had some form of autonomy or intentionality. Interestingly, the authors note that one of the first insights from the design exploration was that the hand-held drones worked with were more fragile than expected -- prone to damage by the tiniest crash -- and required constant care and input. The authors go on to describe the competitors, \textit{"reconstituted the drone from being something simply fragile into something ’vulnerable’"}, and how their design processes actively co-produced vulnerability and risk in the interactions between the humans -- with designerly aims and fleshy bodies -- and the drone technology with its technological affordances and fragile plastic body. 
Through interactional work within the assemblage of humans and drone technology, Popova et al.~\cite{popova2022vulnerability} opened up a design space in which to engage with an unfamiliar technology, non-habitual design activities, and exploratory ideas. Their account of an exploratory design process that unfolded over a significant period of time provides an interesting contrast to the drone challenge we focus on in this paper, in that the challenge participants were brought to explore and develop human–drone interactions within the limited time frame of a three-day event, without dedicated training in interaction design or somaesthetics, and toward the specific objective of piloting the drone through a competitive obstacle course.

\section{Drone Challenge}

The drone challenge was set up as a three-day hackathon. It took place in a unique space [anonymised for peer review] on a university campus in [anonymised for peer review].

\subsection{Objective and Rules}
\label{sec:objective}

\begin{figure}[tb]
  \centering
  \includegraphics[height=3.5cm, keepaspectratio]{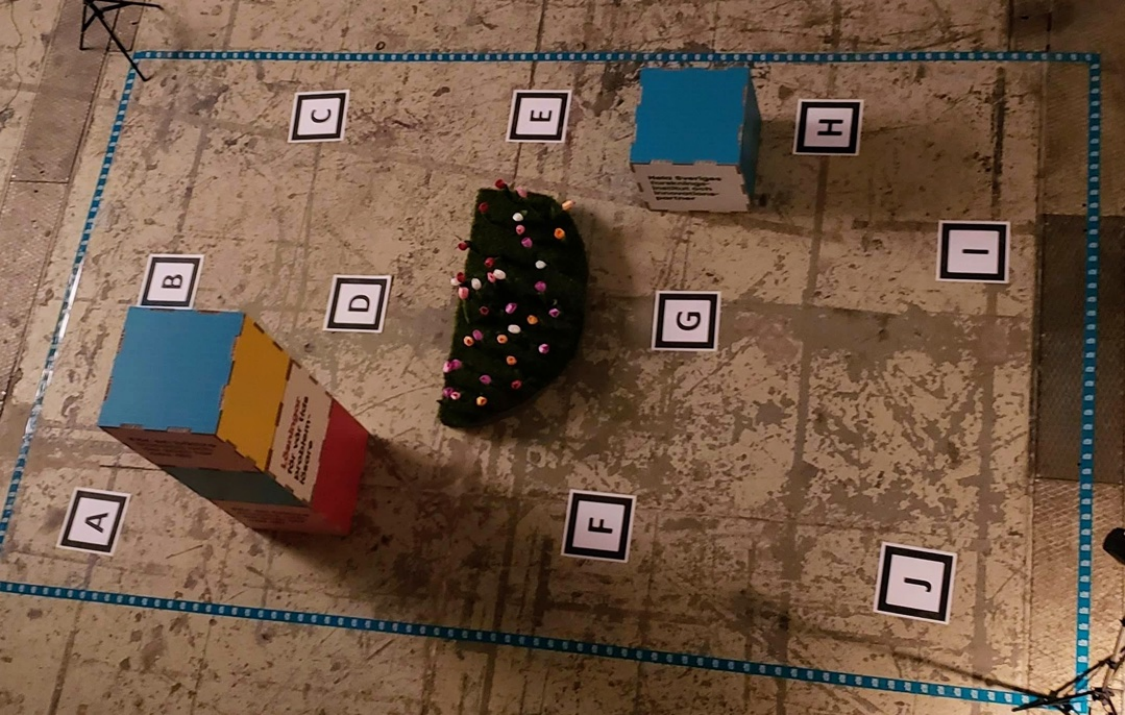}
  \vspace{-2mm}
  \caption{One of the drone arenas, seen from above.}
  \vspace{-4mm}
  \label{fig:arena}
\end{figure}

%\begin{wrapfigure}{R}{0.32\textwidth}
%\centering
%\includegraphics[width=0.28\textwidth]{pics/gestures.pdf}
%\vspace{-2mm}\caption{\label{fig:gestures}A participant steering the drone with hand gestures.}
%\vspace{-4mm}
%\end{wrapfigure}

We designed the competitive challenge taking inspiration from the '80s maze action videogame Pacman.  
The goal was to fly a \emph{nanodrone} through an obstacle course set in a \emph{drone arena}, exemplified in \figref{fig:arena}, in a fixed time while collecting as many markers as possible.
Drone piloting was achieved by a competitor-participant who physically interacted with the drone in the arena, for example, by making gestures that were detected by the drone's onboard \emph{proximity sensors}. Collecting a marker is achieved by successfully flying the drone over a it at any height.

The team members implemented the drone control logic running on a nearby control station as a simple Python program. 
The program determined how a drone would react to a human pilot's interactions. For example, it steers the drone sideways when a participant moves their hand closer to the drone on one side. 
This is a form of \emph{assisted} drone piloting achieved by physically interacting with the drone.
Only one human pilot at a time could be active in the arena during a run.
Participants could not touch the drone or install anything in the arena, such as additional markers or sensors.

We purposefully set up the event so as to discourage participants from taking an unnecessarily competitive or rushed approach to the challenge. In introducing the event to the participants, we stated learning about the drones and having fun as the main goals of the challenge rather than winning, although we did explain that there will be prizes for the most successful teams. Framing the challenge around learning and emphasising that the participants were free to decide how much time they wanted to spend working on the challenge, we tried to establish a calm and supportive atmosphere. We did this to center values like curiosity and collaboration, rather than aggressive rivalry and competitiveness.

\subsection{Technology}
\label{sec:tech}

\begin{figure}[tb]
  \centering
  \includegraphics[height=2.8cm, keepaspectratio]{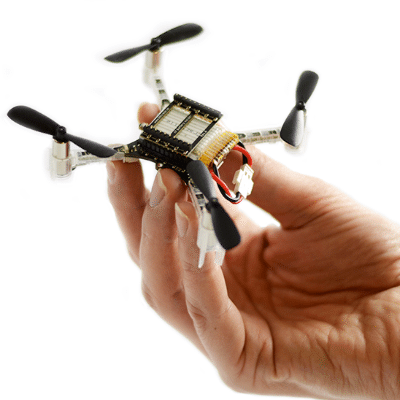}
%  \hspace{5mm}
%  \includegraphics[height=4.2cm, keepaspectratio]{pics/cfclient.png} 
%  \vspace{10mm}
  \includegraphics[height=2.8cm, keepaspectratio]{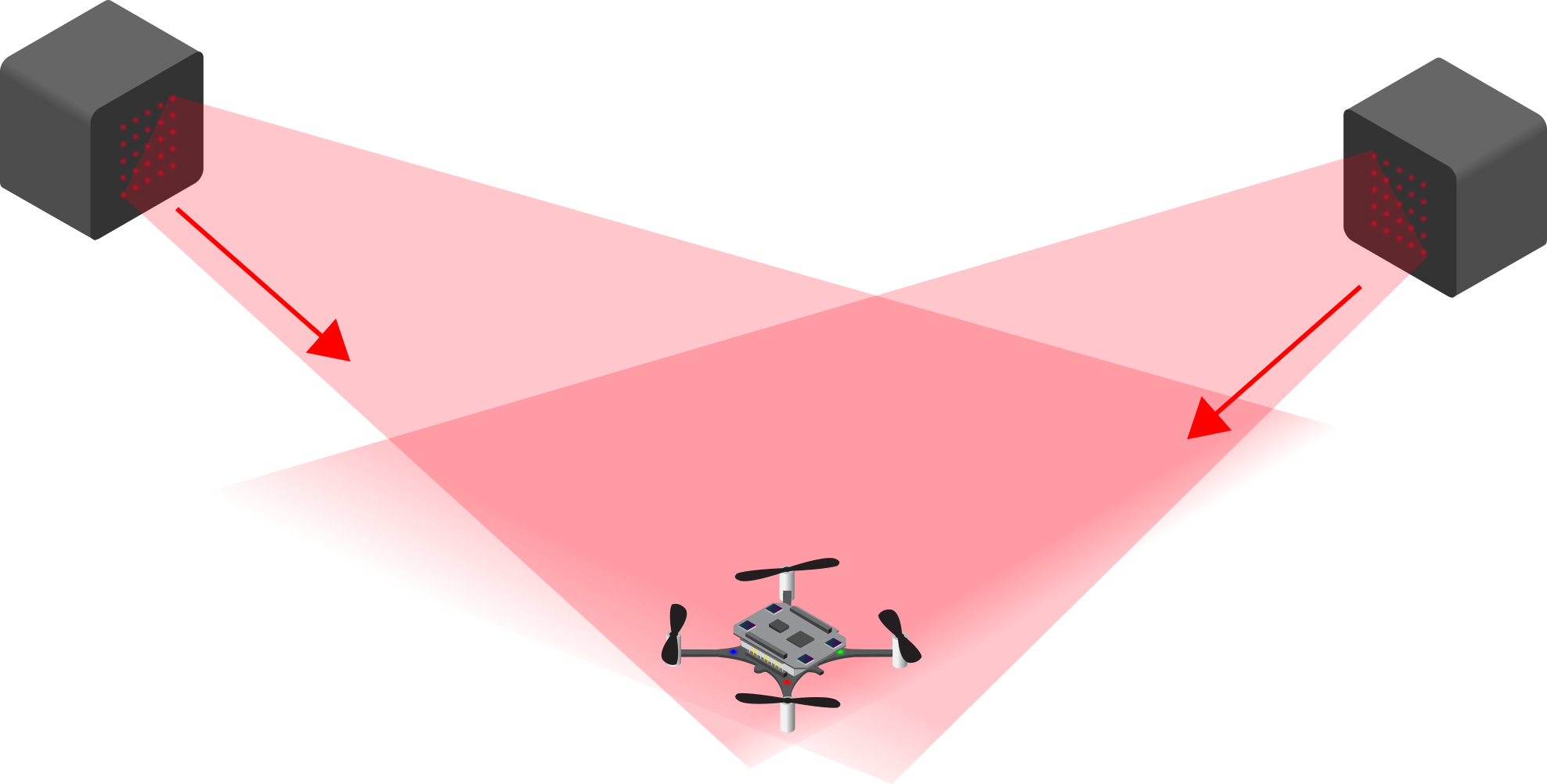} 
   \vspace{-2mm}
  \caption{Crazyflie nano-drone.}
  \label{fig:cf}
  \vspace{-3mm}
\end{figure}

We used the \cf 2.1, shown in \figref{fig:cf}, an open source nano-drone platform that only weighs 27 g and fits in the palm of a hand.
The \cf offers a complete development environment including a piloting GUI, a rich set of application programming interfaces, detailed documentation, and tutorials.
We also provided examples of the control logic running on the control station that showed participants how to achieve simple piloting functionality, such us "pushing" the drone in a given direction when the drone's onboard sensor detects that someone's hands are approaching, or "pulling" the drone when the hands are withdrawing.

An essential element to achieve this functionality is localizing the drone in a 3D reference system.
As the challenge took place in an indoor setting, competitors were unable to rely on GPS and the like for this.
Instead, we equipped the drone arena with the Lighthouse localization system.
This system used two base stations deployed at opposite corners of the arena to emit infrared laser scans, as depicted at the bottom of \figref{fig:cf}. These were detected by the drone using dedicated sensors. Based on the difference in time of arrival between the scans from different base stations, the drone technology could estimate its position. The Lighthouse localization system was sufficiently simple to install in a temporary location and provided a user experience largely similar to other indoor localization systems, including optical ones, such as OptiTrack~\cite{afanasov2019flyzone}. 

Needless to say, the Lighthouse localization system worked reliably only as long as the laser scans were able to constantly reach the drone, similar to an OptiTrack system requiring line of sight to the markers aboard the drone. If the path from a base station to the drone was somehow occluded, say because a person moves in between the two, the drone temporarily lost the base station inputs. Depending on the number of base stations occlusions and on the duration of the disruption, the drone temporarily became unstable yet eventually reclaimed a stable behavior, or completely lost positioning information resulting in a crash.
The operation of the localization system was a significant emergent factor in shaping relationships between pilots and drones during the challenge.

\subsection{Schedule and Teams}

The challenge began with a kick-off event at lunch time, including an \emph{opening talk} intended to inspire the participants, as well as a  \textit{a two-hour tutorial} on programming the \cf, using teaching material developed specifically for the event. By the end of the tutorial, even teams with no previous experience with the technology were able to fly their assigned drone. 
The remainder of day one and the second day were devoted to \textit{challenge trials}; teams were free to work shorter or longer hours as they preferred, to program their drone to interact with their nominated pilot in the arena. 

We set up two separate drone arenas for the trials, experts on our team were available to provide technical support and spent much of their time helping the teams resolve technical issues before they  submitted their code for the final challenge the following day. 
Day three of the event began with the obstacle challenge, observing  the objectives and rules outlined in \secref{sec:objective}. Each team was allowed four attempts to complete the obstacle course. 
The event concluded with with a short prize ceremony and lunch.

Participation was open to anyone irrespective of age, profession and past experience with drones, on a first-come-first-served basis. Participation was free of charge and, when circulating the call for participation, we advertised prizes for the most successful teams, including wireless headphones, movie tickets, and the opportunity to keep the drone the team had worked with. As we provided all necessary drone hardware, a team only needed to bring a computer for programming. While the challenge was advertised as open to all, the details of the challenge likely made it more compelling to those with some familiarity with drones and/or at least some programming skills. The timing of the challenge during three weekdays in June, along with its location on a university campus, also likely played a role in making it particularly attractive to students.

Six teams signed up and five participated fully in the challenge, for a total of 22 participants. 
All teams but one were entirely composed of university students.
Most participants had previous programming experience in a variety of languages, including Python, but only a few had prior experience of drones, e.g. in activities like aerial filmography that had little overlap with the challenge.
%We further scrutinize each team in \secref{sec:teams} when reporting on their developing drone relationship.

\section{Material and methods}

Next to running the drone challenge as a demonstration of movement-based drone piloting and an opportunity for participants to learn about drones, we approached it as an opportunity to study HDI. %We describe the diverse data collection we conducted to document the challenge as well as the subsequent data analysis.

%\subsection{Data Collection}

Our \textbf{data collection} encompassed multiple methods. First, we took \textit{field notes}, detailing our observations and experiences. These include notes from both the researchers who served as technical support -- and hence heard first-hand about issues the teams were facing and the solutions they were resorting to -- and the rest of the team whose main task was to observe and record the event, drawing upon their training in the social sciences and human–computer interaction. Second, the teams needed to submit \textit{the code} they produced for controlling the drone ahead of the challenge. This gives us a detailed view into the design choices they made and provides evidence of some of the technical options that were explored but ultimately abandoned. Third, we \textit{video recorded} large parts of the challenge. We used stationary cameras to capture the activity in each of the arenas during the trials and the challenge. We used smartphones to capture shorter instances of teams' activities and interactions by their desks and in the event space. Fourth, two researchers conducted short \textit{group interviews} with the teams, asking about their backgrounds, their motivations for participating, and their experiences of the challenge. These complement the insights into the teams' experiences we gained through observation.

We conducted our \textbf{data analysis} in the spirit of \textit{bricolage}~\cite{rogers2012contextualizing}. The code and the field notes, taken together, provide a rich description of the challenge. We worked collaboratively to reflect on observations from revisiting field notes and inspecting the code ~\cite{fagan2002design}, scrutinizing overlaps and distinctions in what caught different researchers' attention during initial data analysis. Through this, we identified \textit{the mutual shaping of pilots and drones}, that is, how participants adapted the drones to match their aims but also how the participants adapted themselves so as to interact with the drones, as a core theme for closer analytic development in which we included also the interview and video data.

We attended to \textbf{ethical research practice} by following standard procedures for informed consent and through conversation among the authors\footnote{The study does not fall within the purview of ethical review in the country where the authors work.}. All participants were given an information sheet as they first arrived to the event. After having had the time to read it and ask any questions, they gave written consent for their participation in the study, with the understanding that the authors would (1) observe the challenge, \mbox{(2) video} record parts of it, and (3) invite participants to take part in interviews, which they could freely choose to join or decline, without any repercussions. We also explained to participants that they could withdraw their participation at any point, without needing to provide a reason, and that we would remove their data from our study if requested. We use pseudonyms when referring to participants and teams.

\section{Findings}
\label{sec:findings}

Our findings are three-pronged. First, we report on how participants' perspectives transformed from the early inspirational imagery of seamless HDI to the reality of fragile, willful drones that are prone to crash. Second, we illustrate participants' bodily interactions to make sense of drone capabilities. Third, we depict teams' explorations in building up their drone interactions, considering the spectrum between shaping drone behaviors to give the pilot more leeway and changing pilots' behaviors to fit with the drone.

%\begin{figure}[tb]
%\centering
%    \includegraphics[width=.9\%linewidth]{pics/keynoteSmall}
 %  \vspace{-2mm}
  % \caption{Opening keynote.}
   %\label{fig:keynote}
%\end{figure}

\subsection{Learning about Drones}
\label{sec:dialogues}

We now illustrate the multiple narratives around drones that emerged during the challenge, how they are reflected in participants' programs, and how they influenced the teams' development of interactions with the drone.

\fakepar{Opening talk and tutorials} The message conveyed during the opening talk was uplifting, evocative, and concentrated on \emph{success stories}. The speaker started from the need to establish a \emph{shared body language} between humans and drones. The imagery shown to the participants was void of any technical detail. 
The speaker drew attention to the challenges in establishing synchrony between people and robots, acknowledging the difficulty of building a shared language: \textit{“When we see people in the movies moving with a robot, with robot bodies that look like our own, it is very easy to imagine what this is about. But what do you do when the robot looks different to you? How do you control, for example, sensing and actuating for additional limbs that have hundreds of joints? The movies made it look really easy but there is a lot that has to happen between your familiar body and this weird robot.”} However, the speaker did not get into the technical details of these interactions. Instead, he demonstrated the development of such shared language through videos showing gestures implicitly commanding drones, akin to what the participants need to accomplish in the challenge. %The most technically-detailed content emerged when, for example, explaining the relative positioning computed by the flying drone compared to two other drones attached to a person's hand, used as positioning devices.
Not a single crash or malfunction was shown at any point.

%\begin{wrapfigure}{R}{0.35\textwidth}
\begin{figure}[tb]
\centering
    \includegraphics[width=.95\linewidth]{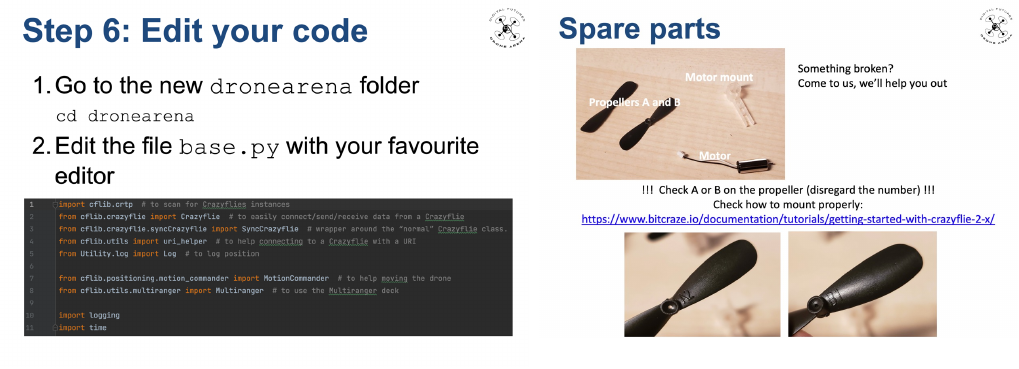}
   \vspace{-4mm}
   \caption{Programming tutorial.}
   \label{fig:tutorial}
   \vspace{-4mm}
\end{figure}
%\end{wrapfigure}

This was in striking contrast to the subsequent tutorial that focused almost exclusively on \emph{technical details}, as in \figref{fig:tutorial}, and was centered on possible issues and how to avoid them. Even the simplest operation, such as getting the drone to take off, was narrated as an activity that requires careful preparations. These include both mechanical aspects, such as assembling the drone in the right way, balancing it manually, and checking it meticulously before turning it on, and actions related to software, since the development environment must be set up correctly and connected wirelessly to the drone, before issuing any command. Roughly half of the time during the tutorial went into describing what could go wrong and how to work around that. The speakers conveyed lots of "hands-on" experience of the sort that is rarely available in manuals.

\begin{figure}[tb]
\centering
    \includegraphics[width=.85\linewidth]{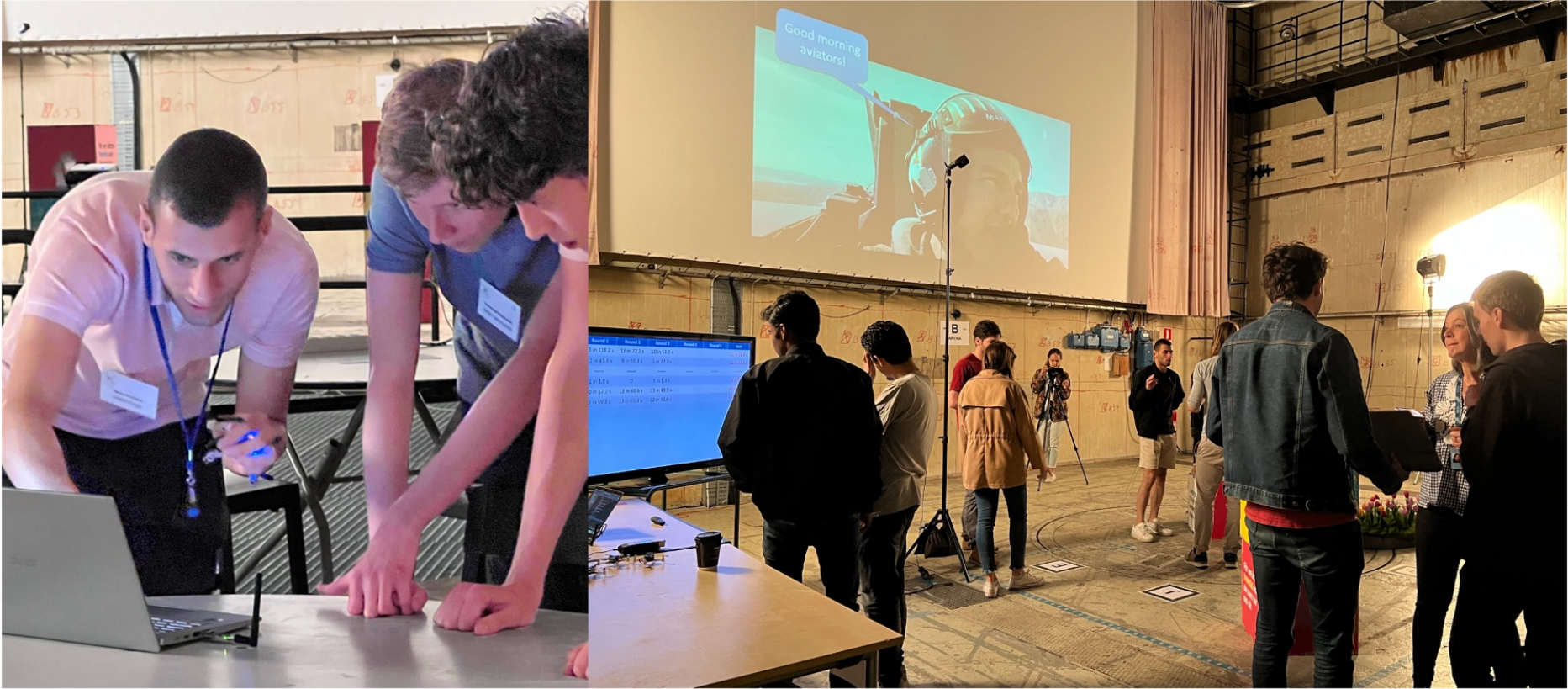}
   \vspace{-2mm}
   \caption{Discussions during the trials.}
   \vspace{-4mm}
   \label{fig:talks}
\end{figure}

\fakepar{Interactions during trials} Discussions during the trials, shown in \figref{fig:talks}, took place between the participating teams and our technical support. They were mostly related to questions about how to understand drone behaviors in the (many) cases when the drone did not behave as expected, the reasons why the drone (most often) crashed, and how to repair it when something broke. Participants came to ask questions like: 
\begin{quote}
\emph{How come that the drone was hovering stably for several seconds and suddenly lost control?}
\end{quote}

\begin{quote}
\emph{Even without imparting any command, the drone keeps moving around... why?}
\end{quote}

\begin{quote}
\emph{How can I see at all what are the readings of the onboard sensors to check what the drone is doing?}
\end{quote}

\noindent The tutors' responses acknowledged the high degree of unpredictability common for drones:

\begin{quote}
\emph{"There may be different reasons why it doesn't take off. Sometimes the best solution is to turn it on and off."}
\end{quote}

\begin{quote}
\emph{"It is really super random. Sometimes it could work, sometimes it doesn't."}
\end{quote}

These situations are arguably an instance of the \emph{visibility problem}~\cite{lee2016introduction}, common in embedded system programming. 
Resource-con\-strained embedded systems, such as drones, often do not run full-fledged operating systems and hence lack most of the inspectio and debugging facilities.
The \cf adopts a form of assisted flight, adescribed in \secref{sec:background}, and the flight controller aboard the drone runs on bare hardware. 
Understanding the execution of programs on a device that has no operating system and only a few LEDs for debugging is often compared to \textit{"looking at an elephant through a keyhole"}~\cite{lee2016introduction}.
Tools exist to address this problem~\cite{wallace1989software} but they are hard to approach for beginners.

During the challenge, the issues experienced by our participants arose mainly due to two factors. First, the sensors aboard the drone used to detect human gestures are both slow and imprecise. The time it takes to detect a nearby obstacle may reach up to a few seconds, defeating the illusion that a mutual relationship between drone and human may develop with the same time dynamics as it occurs between people. 
Second, the localization system, described in \secref{sec:tech}, requires constant line of sight between the base stations and the drone, or the latter loses control and crashes. 
The participants did not immediately realize this. 

As the instructors progressively answered the questions and explained the reasons why these situations occurred, participants entered a mutual adaptation loop, adapting the programs piloting the drones and/or their own body movements, depending on what factor they considered easier to handle.
The transition from the early representation of smooth and expressive HDI in the opening talk, to brittle behaviors and crashes in the arena was abrupt. This shift in perspective is apparent also in the participants' programs.

\fakepar{Programs} We apply established code inspection techniques~\cite{fagan2002design} to examine how the different narratives around drones manifest in participants' programs. As concrete evidence, \appref{app:codeTeam2} reports an excerpt of the program by the team named \textit{Bellman Brothers}.

Many teams began with high ambition, aspiring to the forms of HDI shown during the opening talk. Those interactions were rich and smooth, yet what was not apparent in the opening talk were a number of technical features that made the input to the flying drone much more accurate than the on-board navigation sensors that we asked our participants to use. Participants came to realize this discrepancy the hard way. 

Through multiple crashes and dialogue with the tutors, their attention turned to dealing with the inaccuracy of the onboard sensors. This is where they started commenting out portions of code which were originally meant to replicate the human-drone dynamics seen in the opening talk. This can be seen, for example, in the existence of multiple versions of the function \code{get\_speed\_command} in \textit{Bellman Brothers'} program (\appref{app:codeTeam2}), or in their work on the \code{which\_region} function, which is defined, implemented, and ultimately \emph{never used}. 
These functions were eventually considered not suited to deal with the unreliable inputs and, thus dismissed.

The process of shaping drone behavior based on its actual capabilities was taken to extremes and entire functionality typically used on drones was intentionally excluded from the flight control loop, such as the processing of inputs from the altitude sensor or from the flow sensor.
The code in the \code{main} function of \appref{app:codeTeam2} used a workaround to exclude both sensors from the processing of the extended Kalmann filter used for flight control~\cite{roumeliotis2000bayesian}. 
These sensors are normally used to achieve better flight stability. 
For the challenge, however, four teams out of five decided to use these sensors to command the drone in the vertical direction using their gestures, and eventually gave up on doubling these inputs for the flight control loop, considering this to be too complex to achieve. 

Participants looked for other approaches that offered greater simplicity and robustness against the volatility and inaccuracies of the onboard sensors. 
From pretending that distance readings are mm-accurate, for example, the teams realized that the ranges they should work with are rather tens of cm. 
The final program of \appref{app:codeTeam2} determines drone reactions to the pilot's gestures based on multiple hard-wired parameters, likely obtained through trial and error. 
We discuss this process in \secref{sec:interactions}. These observations apply to the programs of three out of the five participating teams, namely \emph{Bellman Brothers, Flying Ferrets,} and \emph{RoboNerds}. Rather than shaping the drone behavior programmatically, two teams, \emph{Robo Geeks} and \emph{Spark Speed}, chose to focus on shaping their own bodily movements. We return to this aspect in \secref{sec:spectrum}.

\subsection{Exploring Drone Capabilities}
\label{sec:interactions}

\begin{figure}[tb]
\centering
    \includegraphics[width=.85\linewidth]{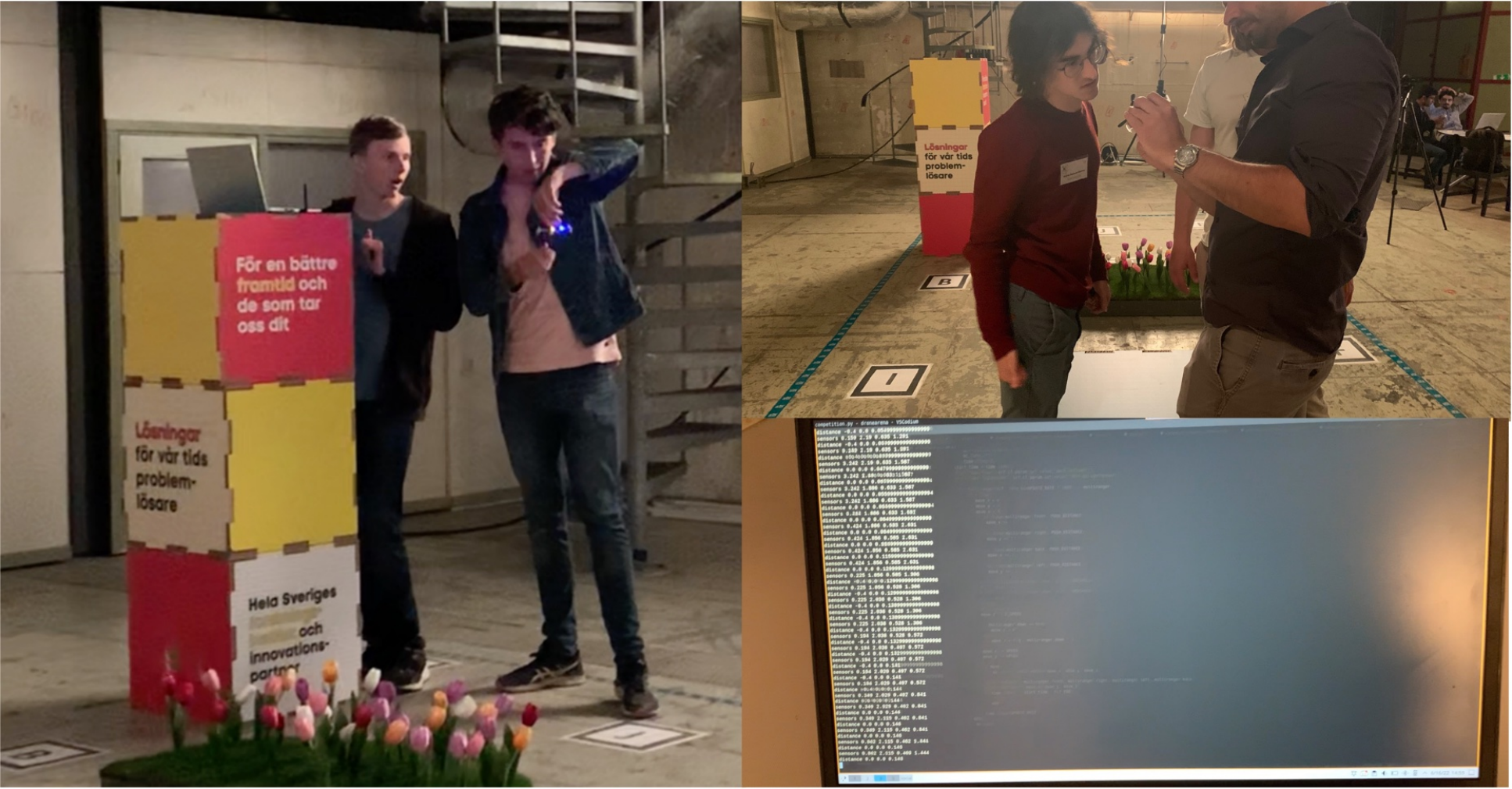}
   \vspace{-2mm}
   \caption{Participants probing the drone to find out its capabilities and limits.}
   \vspace{-4mm}
   \label{fig:interactions}
\end{figure}

Once the participants realized that what was in the opening talk was not attainable out of the box, they started a process of exploring the drone's capabilities to find out what actually was feasible. This spanned both the drone's hardware and software. As one participant described, this involved collaborative experimentation with different parameters where one team member was in charge of tweaking the parameters by the computer while another interacted with the drone: \textit{"A big part of the problem is just tuning the parameters: the velocity, tuning the push distance, and so on... So we have to try and see is it too fast now? Is it too slow? Is it... Do we need to move too close to it to make it react? So we just try different parameters and see how well it works and our rules is, we don't have the rules, but Leif deals mostly with the computer, he changed the parameters. I'm mostly with the drone, setting it up and moving with it."}

Throughout the event, teams were constantly experimenting in the arenas: changing one parameter at a time and verifying the effect of those changes on the drone behavior. 
This required plenty of touching and bodily adjustments to the drone, checking connections among the different components, understanding the weak points and what parts are prone to break, plus fixing whatever broke. 
We observed participants tilting drones, trying to identify where is top and bottom, and sometimes hurting their fingers with propellers in the process. %In interviews, participants brought up feelings of fear, both for their own safety and for the drone: \textit{“When the drone comes, comes to you from this part [pointing at the torso] I feel more vulnerable, but when it is around a foot, I am not as much vulnerable. When it is around a head or the eye, I feel like I am in danger. But when it is just going around on the ground, I am just worried that the drone itself gets damaged so they cannot fix it"}. 
The participants also reported on learning to adjust to the drone and developing more and more courage in handling these fragile devices: \textit{“They always look so fragile. So first you are afraid to just handle them: whether you will just destroy it or a sensor will come out. So you have to handle them with a lot of care.”} These feelings became essential to participants' developing understandings of the drone capabilities, even though they are not commonly mentioned in papers or drone manuals. 

Some teams eventually created \textit{boundary situations} to push the drone to its limits and determine where those are.
One team admitted to have tried out an approach whereby the drone coped with obstacles by physically bumping into them, making the sensing problem essentially irrelevant as the drone would directly feel the obstacle when it touches it. The team quickly abandoned this idea, facing irremediable breakages at every encounter with an obstacle. In general, crashes became the norm rather than the exception; participants appeared surprised and felt amused whenever things actually (momentarily) worked. Some participants performed several \emph{mock-up flights}. As shown on the left of \figref{fig:interactions}, team \emph{RoboNerds} entered the arena with the drone in their hands and mimicked an actual run. 
While doing so, another team member was monitoring on a laptop the sensor readings coming from the drone as it was approaching obstacles or being subject to different piloting gestures.
The data was instrumental to understand the accuracy of sensors and tune the different parameters in the code.

These procedures may be regarded as a crude, bodily approach at solving the visibility problem. 
Rather than relying on dedicated industry-strength solutions, participants engaged in an intimate form of bodily interaction with the drone to \textit{"get to know it"}, as one of the participants in \emph{Flying Ferrets} put it. These activities -- not required by the challenge but ultimately productive for achieving its objectives -- unfolded over time, likely because the teams were gaining deeper and deeper understandings of the drone and eventually \textit{"got to know it"}, as the same participant declared at the verge of the final day, feeling confident of their outcome.

As the trials progressed and especially during the actual challenge, some participants started interacting with the drone as if it was a \emph{living entity}, capable of communication on a level more advanced  than they are capable of. 
We observed participants talking to the drone, suggesting where to move and making gestures akin to giving directions to another person rather than to a robot. While such activities could, of course, be interpreted as directed toward the other people observing the pilot's interaction with the drone -- making one's actions and struggles accountable to a human audience -- here they came across as genuine attempts at HDI.

\subsection{Training the Drone vs.\ Training the Pilot}
\label{sec:spectrum}

We observed strikingly different approaches at shaping the HDI necessary to tackle the challenge: 

\fakepar{Training the drone} Some teams largely concentrated their efforts on producing a program that was sufficiently intelligent to handle sensor inaccuracies and could accurately respond to the pilot gestures. \emph{Bellman Brothers'} code (\appref{app:codeTeam2}) is one such example, but this approach was taken to an extreme by another team, the \emph{Flying Ferrets}, whose program we report in \appref{app:codeTeam5}. This team demonstrates how background and prior skills impact the shaping of HDI: members of \emph{Flying Ferrets} were enrolled in an Aerospace Engineering program and had solid knowledge in flight control, including the use of Kalman filters~\cite{welch1995introduction} for state estimation.
%A Kalman filter is an algorithm that uses a series of measurements observed over time, possibly affected by noise and other inaccuracies, and produces estimates of unknown variables, such as the distance from an obstacle in this case, that tend to be more accurate than those based on a single measurement alone.

\emph{Flying Ferrets} was the only team who wrote their code from scratch, instead of adapting one of the examples provided during the tutorial. 
Their parameter setting and velocity estimations, as implemented in function \code{compute\_velocity}, demonstrate their intimate knowledge of where and how input data is gathered and of the sensitivity of velocity estimations. The team explicitly defined parameters to strike a trade-off between maximum velocity and accuracy of control in their code, as in \code{ActionLimit} and \code{VelocityLimit}. \emph{Flying Ferrets} were also an example of \emph{defensive programming}~\cite{qie2002defensive}, which they used to handle sensor inaccuracies.
Defensive programming is the creation of programs designed to avoid problematic issues before they arise. 
One common method for achieving this is through code that is meant to deal with any possible scenario thrown at it, making the program able to run properly even through unforeseen situations. 

Defensive programming often relies on considering situations whose concrete occurrence might not be ever verified. \footnote{Existing statistics~\cite{cote1988software} report that, when adopting this form of programming style, roughly 30\% of the produced code is never actually executed in production.
In many ways, the concept is akin to that of defensive driving, in that problems are considered before they arise and not because they occurred in the past.}
We found evidence of defensive programming, for example, in the change of the default reading frequency of sensors, which many teams applied up to a $5x$ increase compared to the factory configuration.  
Obtaining more data allowed these teams to apply aggressive averaging and filtering to exclude outliers.
Team \emph{RoboNerds} implemented buffers of up to 100 readings to do so, which is 10 times more than what is normally performed on most existing flight controllers~\cite{bregu2016reactive}.
This increases the drone stability during flights, in exchange of slightly higher energy consumption.
The latter, however, was not an issue during the challenge, as the battery lifetime far exceeded the duration of a single run and the technical support made plenty of spare batteries at disposal.
Most teams also capped the maximum drone velocity down to half of what the \cf can do, in an attempt to make things easier for the human pilot.

\begin{figure}[tb]
\centering
    \includegraphics[width=.8\linewidth]{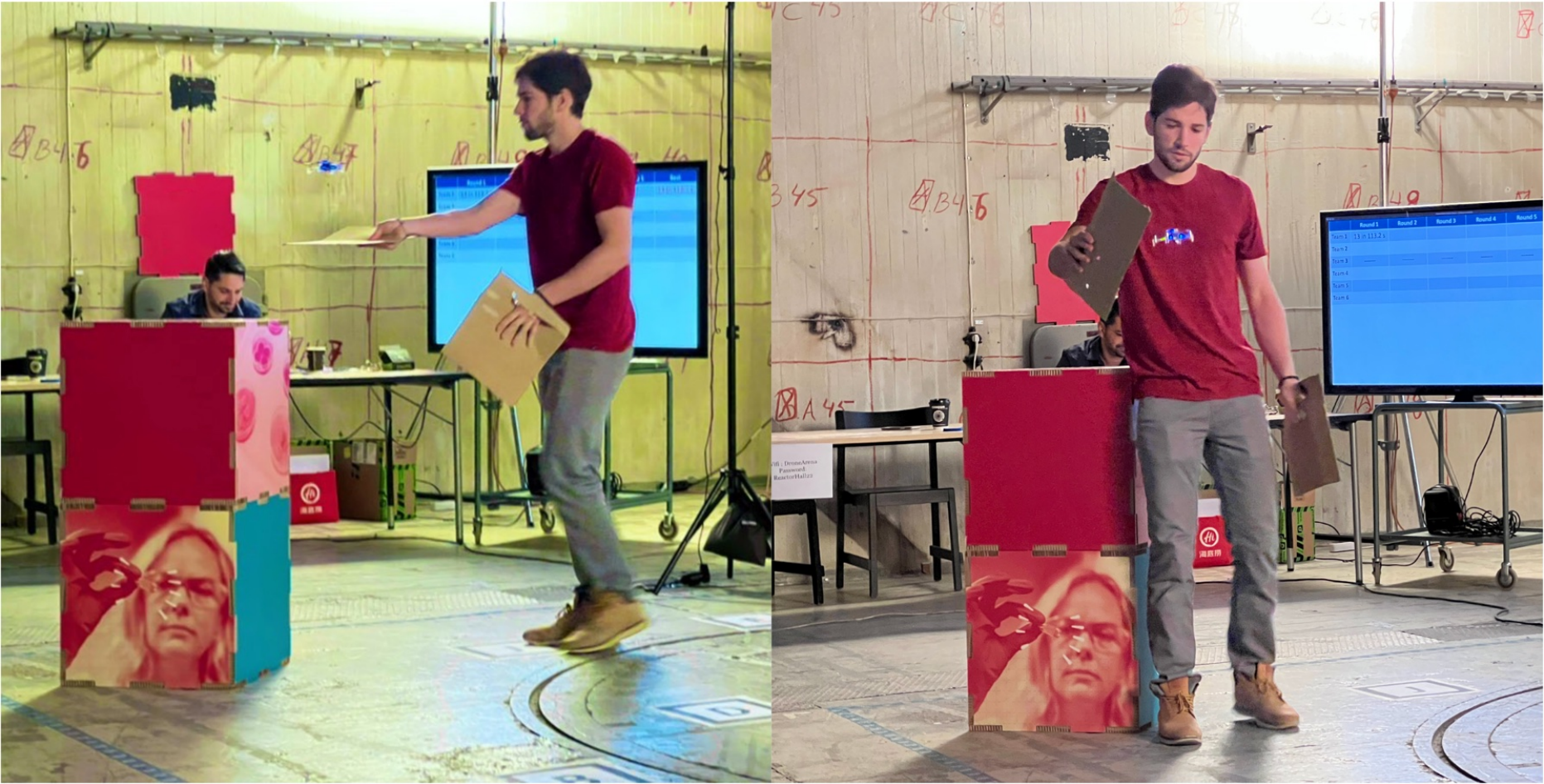}
   \vspace{-2mm}
   \caption{A participant using cardboard paddles as an extension of his body to pilot the drone.}
      \vspace{-3mm}
   \label{fig:paddles}
\end{figure}

\fakepar{Training the pilot} At the opposite extreme, we find teams that applied very limited changes to the examples provided during the tutorials and rather spent most time in the arenas in an attempt to gain the necessary piloting skills. 
One team, \textit{Robo Geeks}, only added 7 lines of code to one of the tutorial examples and spent most time in the arenas running that program, learning how to best command the drone with their gestures.

According to our recordings of the event, \textit{Robo Geeks} spent roughly 40\% more time in the arenas than any other team. 
During the actual challenge, \textit{Robo Geeks} were inadvertently penalized for this approach -- which held a lot of promise throughout the trials -- because none of the code examples we provided was sufficient per sae to reach all markers. 
The marker placed on one of the obstacles, in particular, required changing flight altitude, which was not possible without tweaking example code. 
\textit{Robo Geeks} recognized the problem on the last day and, after initial surprise and upset, decided to simply skip the problematic marker during the challenge, hoping their well-rehearsed piloting would compensate for this shortcoming by collect all other markers more quickly.

\begin{figure}[tb]
%\begin{wrapfigure}{R}{0.3\textwidth}
\centering
\includegraphics[width=0.25\textwidth]{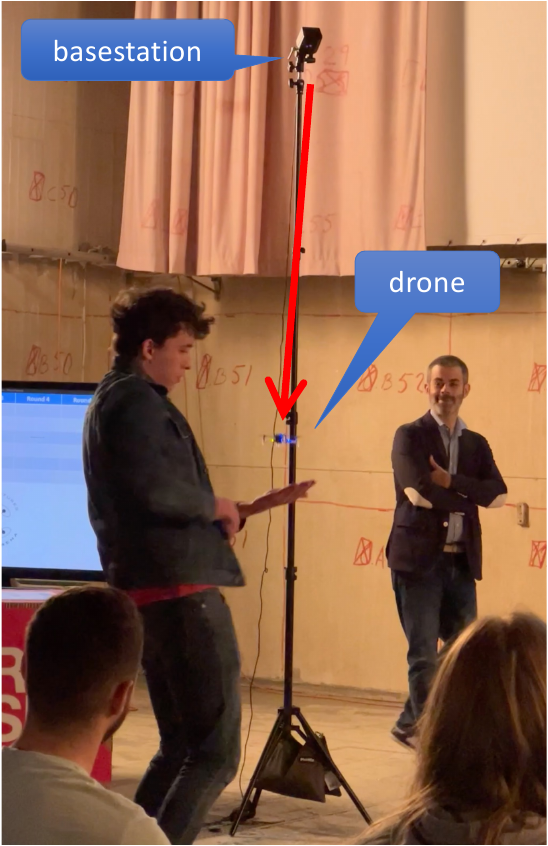}
      \vspace{-2mm}
\caption{\label{fig:beam}Stilted movements to maintain line of sight between Lighthouse basestation and drone.}
      \vspace{-3mm}
\end{figure}
%\end{wrapfigure}

As a result of shaping their own movements, some teams eventually discovered new and unconventional ways to pilot the drone. 
Team \textit{Spark Speed}, for example, realized that in addition to the two inputs provided with their hands, further input may be provided by other body parts, especially the belly.
This allowed them to better control the drone as it could be piloted through one additional input.
Another team, \textit{Cyber Ravens}, came to think that their own hands were one source of sensor inaccuracies. This has some technical foundation: the beam of the drone sensors is narrow, so extending the sensed surface may provide more accurate and stable readings.
To extend the pilot's physical body so as to provide better inputs to the drone, \textit{Cyber Ravens} eventually opted for wearing self-made cardboard paddles to command the drone, as shown in \figref{fig:paddles}, to extend the area that the drone sensor could possibly detect. 

Most pilots also performed atypical, stilted movements to cope with the limitations of the localization system, which requires line of sight between the base stations and the drone. Some of these, such as the example in \figref{fig:beam}, put significant strain on the pilot's body. We observed participants bending their backs to stay in the vicinity of the drone with their hands and provide continuous piloting input without standing in the way of the localization system. During the trials, one participant attempted piloting the drone on his knees, but this turned out too impractical and unsafe as the drone propellers ended up on level with the participant's eyes.   

In the end, team \textit{Spark Speed} won the challenge, striking the best trade-off between shaping the drone behavior and extending their own capabilities by using not just their hands, but other parts of the body as piloting input. Team \textit{Cyber Ravens} ranked second, thanks in part to their use of cardboard paddles as an extension of the pilot's hands. \textit{Flying Ferrets}, with their accurate tuning of state estimation algorithms, ranked third. %Remarkably, all teams were able to complete the obstacle course within the set time in more than 50\% of the runs during the actual challenge.

\section{Somatic engagement in the design and development of drones}

We acknowledge that the drone challenge is an edge case in that we look at a group of people who are given a limited amount of time to get to know the drone and shape their interaction with it. %Yet, most participants had relevant technical skills and all had access to competent tech support that helped compensate for the lack of time to get to know the technology. 
Our call for action, here, is based not only on participants' experiences, but also the tech support and experts on drone technology in our team. As we argue next, getting to know the drones is a deeper and more involved process than merely learning to understand how to program them, or figuring out their sensor system. 

Messy details of interacting with the drone were at the core of the challenge. The team work mostly consisted of failures: crashes and breakdowns that led to better understanding of the drone, fostering \textit{bodily knowledge} on how to handle the drone, how to optimise the code, or how to make sense of the sensor readings. Handling failures took most of the three days of the challenge. Only the final runs during the challenge itself went (for the most part) smoothly. This said, even the final performances of the winning team were at times interrupted by technical problems. These constant problems and consequences of breakdowns are in stark contrast to the \textit{narrative of success}~\cite{howell_cracks_2021} that we often see in design reports. However, as we argue here and is highlighted in prior work, not only are real-world robots far from perfect~\cite{10.1145/3461778.3462060} but breakdowns are, in fact, essential for design practice as they can help us develop technical qualities, create mutual understanding within a design team, and let the team explore and exploit the affordances of the technology at hand~\cite{popova2022vulnerability} -- in this case, understanding the complexities of drones. %We, therefore, echo calls~\cite{howell_cracks_2021} to refrain from binary dichotomies between successes and failures.

Considering teams' explorations to get to know the drones, we note that the challenge involved not so much coding  -- although some teams chose to write their own -- as tweaking parameters: changes were often not algorithmic but adjustments to improve how the drone's sensing picked up on the pilot's movements. The challenge, then, could be framed as changing, rather than shaping, the drone. Importantly, we have illustrated numerous situations where what is trained and shaped is not the drone but the pilot.
%As reported above, one of the teams decided to focus almost entirely on training their pilot, adjusting their program only minimally. This approach of focusing on training the pilot held a lot of promise throughout the trials, even though it ultimately introduced some challenges, in particular when the obstacle course was changed ahead of the challenge runs in a way that would have necessitated the drone to have capabilities that it had not been coded to have.
We also saw many instances of pilots making bodily movements to figure out how the drone would respond; while the pilot was moving with the drone, another team member looked at the screen, trying to follow what data is generated and when exactly something goes wrong. It is, after all, impossible to move with the drone while also looking at the screen. 

This begs the question if we could have given the teams better tools to explore the drone behavior?  
There exist a fundamental gap between coding the drone behavior and its behavior once the program runs. 
The issue largely stems from the \textit{visibility problem}~\cite{lee2016introduction} we discussed in \secref{sec:dialogues}: much happens "inside" the drone that is not visible from the "outside". 
Since embedded programming environments for drones are still rather rudimentary, the drones often became black boxes, producing incomprehensible behaviors. While we could do our best to create better embedded programming environments~\cite{mottola2006pervasive} -- simulators, visualisations, proper debugging and ways of watching live data streams  -- these issues also point to an underlying problem relating to \textit{somatics and bodily knowledge}. 

As Suchman~\cite{suchman1997interactions} argues, to act intelligently in a situation, machines need to become \textit{readable/writable}. 
A pre-requisite for this is that their inner states must become \emph{visibile}. 
Machines must convey their inner state, making them \textit{readable}, but they should also provide us with \textit{inscribable surfaces} that allow us to dynamically change our plans and behaviors to attain our goals depending on, and in response to, what the machine does \cite{suchman1997interactions}. We acknowledge that achieving this within the limited computing and energy envelop of nano-drones, usually equipped with low-power resource-constrained microcontrollers, is incredibly challenging. Yet, we argue that to understand drones, we can focus solely neither on watching their behaviour, nor figuring out how they sense human movement. We need to enter a \textit{perception-action loop} where we can both probe and change the drone's behaviors. 

To support \textit{perception-action loops}, we can draw on the evolution of personal computing and an analogy to direct-manipulation. As discussed by Mueller~\cite{mueller} \textit{"computing started with machines that executed a program in one go before returning the result to the user. By decreasing the interaction unit to single requests, turn-taking systems such as the command line evolved, which provided users with feedback after every input. Finally, with the introduction of direct-manipulation interfaces, users continuously interacted with a program receiving feedback about every action in real-time"}. 
%We draw on this analogy to call for ways to directly-manipulate drones' sensing and actuation mechanisms in and through movement, as a way of programming their behaviour.
%This is where we draw on the analogy to Mueller's work on direct-manipulation. 
One of Mueller's systems, FormFab, allows for such direct control of a 3D-printer, specifically, of the vacuum forming process which is typically a high temperature moulding process that only allows for one shape to be explored at a time. %FormFab uses a work piece that, when warmed up, becomes compliant and can be reshaped. To re-shape it, users direct-manipulate a pneumatic system interactively. As users interact, they see the shape in the 3D-printer change in real-time. A similar solution could be applied to drones. As we cannot directly touch a drone without being harmed or breaking it, we instead envision a system that can be direct-manipulated and touched as a way of direct-manipulating the drone. 
FormFab allows the user to select a section of the work piece (a plastic sheet) to be heated by a robotic arm. This section, then, becomes compliant and can be reshaped by manipulating a pneumatic system interactively. As users interact with the system, they see the plastic sheet change in real time as a result of the pressure changes. A similar solution could be applied to drones to leverage the bodily knowledge of the designer in shaping the control systems of the drone. 
%%

% As Norman~\cite{norman2010natural} points out, designing for some "natural movement" will not help as meaning-making arises from seeing the consequences -- the \textit{feedback} -- of a gesture or movement as it unfolds in real-time. This feedback needs to be overtly accessible throughout the whole interaction. Norman \cite{norman2010natural} frames it as: \textit{"More important, gestures lack critical clues deemed essential for successful human-computer interaction. Because gestures are ephemeral, they do not leave behind any record of their path, which means that if one makes a gesture and either gets no response or the wrong response, there is little information available to help understand why. The requisite feedback is lacking"}. That is, programming environments for drones could be of substantial help if they could make those action-feedback loops somatically felt in the moment, as they unfold over time.

Crucially, though, issues with drones do not arise solely from a lack of accurate real-time feedback. The material body of the drone itself is fragile and error-prone. In our challenge, we noted repeatedly how teams were mystified as to why their drone acted the way it did. Was it the battery failing? A broken propeller? A sensor malfunction? Teams had to tease out the fragilities of the drone's physical body from other problems, such as a bug in their program, or anomalies, like when the pilot gets in the way of the positioning technology. Telling the different potential sources of trouble apart was far from trivial. %As mentioned, this problem was exacerbated by the compressed time frame and competitive environment the teams had to work in. %With these external pressures, it is not possible to become familiar with the drones, to the point where the designer can reflect on the aesthetics or efficiency of their movements in relation to the drone \cite{la2020designing}. Nor how to improve its behaviour in order to win the competition. 
The messiness of the design process that the teams experienced does not dismiss but rather supports our call for programming environments that elucidate action-feedback loops for somatic appreciation. 

Movement-based programming systems \cite{fiebrink2010wekinator, interactiveml} are gaining popularity as designers increasingly work with complex algorithms and programming techniques they have not been trained for~\cite{yang2020re, bogdan2020programming}. We see the somatic training of designers who interact with these movement-based systems as part of a multi-faceted approach to the somatic processes and bodily understanding required to utilize these systems to greater effect. We argue that systems purporting to elicit supposedly 'natural' gestures or movements \cite{naturalhdi} often assume that such bodily knowledge does not need to be 'trained' or 'developed' in the same way as technical skills, such as coding. However, our study suggests that the cultivation and development of somatic capabilities also plays a fundamental role when developing movement-based robotic systems.

Similarly as regular programming systems evolved from textual representations of the application logic to graphical ones, our call for action here is to push programming environments for drones and the like to \emph{elucidate perception-action loops} through somatic engagement. %, rather than merely enabling user interaction. 
The aim should be to support people's engagement in these loops rather than forcing them to break out of it which is what happened during the challenge when participants needed to go back-and-forth between checking the code and piloting the drone.
We envision augmenting HDI design processes by making space for somatic sense-making of autonomous robot technologies \cite{ghajargar2021explainable, ghajargar2022graspable} as this would serve designers in their efforts to uncover the various possibilities -- both creative and technological -- offered by such technologies \cite{fdili}. 

%%<and possibly a comparison to.. ta-da! ... HORSES!>

%In addition, they made pilots engage in awkward movements, not making it easy on the human body. 

% body language not just about beauty but also optimising how they move their own bodies

%\subsection{Supporting somatic engagement in the design and development of drones and other autonomous systems}

%We argue for a felt approach to designing and developing autonomous machines, such as drones, and conclude with a discussion of how we might encourage and support designers to work with movement, body and feeling throughout the programming, rather than having to constantly divide their attention between a computer screen and the drone.

%ref back to programming with the body paper

%a human-drone simulator? or should we avoid doing it on the screen and rather do it in the room?

%generalizing to go beyond just drones

\section{Conclusion}

We contribute with an empirical, somaesthetically focused account of current challenges in HDI. Reporting on a three-day drone challenge, we have illustrated how teams \emph{1)} shifted from aiming for seamless HDI, to seeing drones as fragile, wilful, and prone to crashes; \emph{2)} engaged in intimate, bodily interactions to more precisely understand, probe, and delimit their drone's capabilities; \emph{3)} adopted different strategies, emphasising either training the drone or training the pilot. In discussion, we argue for supporting somatic engagement in the design and development of drones with the help of programming environments that elucidate and enable \textit{action-feedback loops}. Embedded programming tools for drones do not only need to allow programmers to adequately change the code or sensor filters of the drones, but also to 'train' their pilots through providing somatic signs and signals that can be probed and felt in real-time~\cite{la2020drone}. This will let pilots learn about how they might need to change their own movements to fly the drone, until they create a new way of moving together with the drone. %The somatic feedback loops we call for should also let pilots disambiguate correct behaviours from  misinterpretations regarding what is going on -- as we have illustrated, it can, at present, be difficult to tell apart the fragilities of the drone's physical body from other problems, such as a bug in program code. 
While aerial drones are in some ways a unique type of technology, we believe that the broader argument regarding the need for supporting somatic engagement in design and development processes is also relevant for HRI, more broadly.
%<something generalising to autonomous systems goes here>
%<and possibly a comparison to.. ta-da! ... HORSES!>

\bibliographystyle{ACM-Reference-Format}
\bibliography{references}

\newpage
~
\newpage

\appendix

\section{Code Example}
\label{app:codeTeam2}

An excerpt of code produced by team \emph{Bellman Brothers}. We use \code{\#...} to omit lines of code that are not relevant for the discussion. Because of formatting, we could not keep the regular Python indentation. The whole program is $\approx$ 400 lines of code.
\begin{lstlisting}
from functools
    import reduce
import time
from cflib.positioning.motion_commander
     import MotionCommander
from Utility.extension.decks.deck
     import DeckType
from utils
     import ActionLimit, VelocityLimit,get_vx, get_vy
from Utility.extension.extended_crazyflie
     import ExtendedCrazyFlie
from Utility.extension.decks.z_ranger
     import ZRanger
import numpy as np 
import matplotlib.pyplot as plt

SPEED_CLIP = 1
BATTERY_LIMIT = 10
ADJUST_VELOCITY = 0.3 # [m/s]
threshold = 0.1 # [m]
DEFAULT_HEIGHT = 1
prev = 0 # 0=hovering, -1=lowering, +1=raising
def which_region(d,push_range,pull_range,unsensible_range): 
    # ...

def adjust_height(zrange_state : float):
    global prev
    h = zrange_state
    if (h < DEFAULT_HEIGHT + threshold) and prev < 1:
        prev = 1
        # zranger.contribute_to_state_estimate = False
        # disable zrange contribution
        return ADJUST_VELOCITY
    elif (h > DEFAULT_HEIGHT - threshold) and prev >-1:
        prev = -1
        zranger.contribute_to_state_estimate = True
        return -ADJUST_VELOCITY    
    elif prev != 0:
        prev = 0 
        return 0  

class state_to_filter:
    def __init__(self,n_steps):
        self.array = np.zeros((n_steps,))
        self.mean_value = 0
        self.measurement_ready = None

    def add_new_value(self,value):
        self.array = np.hstack((value,self.array[:-2]))
        # mask = (self.array != 0.)
        self.mean_value = np.mean(self.array)
        # self.measurement_ready = self.array[0]!=0
n_avg = 20
front_range = state_to_filter(n_avg)
back_range  = state_to_filter(n_avg)
left_range  = state_to_filter(n_avg)
right_range = state_to_filter(n_avg)

#def get_speed_command(distance,lower_bound,upper_bound):
    # ... 

def get_speed_command(distance,lower_bound,upper_bound):
    if distance>lower_bound and distance<upper_bound :
        gap_width = np.abs(upper_bound-lower_bound)
        current_gap_dist = np.abs(distance-lower_bound)
        return 0.4
    else :
        return 0

def script(multiranger_state : dict, mc : MotionCommander) :
    # ...           
          front_range.
             add_new_value(multiranger_state['front']/1000)
          back_range.
             add_new_value( multiranger_state['back'] /1000)
          left_range.
             add_new_value( multiranger_state['right']/1000)
          right_range.
             add_new_value( multiranger_state['left'] /1000)
  
          v_front =
                - get_speed_command(front_range.mean_value,
                                    push_range_start,
                                         push_range_end)
          v_back =
                get_speed_command(back_range.mean_value,
                                    push_range_start,
                                       push_range_end)
          v_right =
                - get_speed_command(right_range.mean_value,
                                    push_range_start,
                                        push_range_end)
          v_left =
                get_speed_command(left_range.mean_value,
                                    push_range_start,
                                       push_range_end)
        # front_readings.append(front_range.mean_value)
        # left_readings .append(left_range .mean_value)
        # right_readings.append(right_range .mean_value)
        # back_readings .append(back_range.mean_value)
          z_range = ecf.coordination_manager.
                get_observable_state(zranger.
                                    observable_name)['zrange']/1000
          #   z_readings.append(z_range)
          vz = adjust_height(z_range)*0
          mc.start_linear_motion(v_front+v_back,
                                     v_right+v_left,vz ) # raise up
    # ...

if __name__ == '__main__':
        # ...
        # disabling height and/or flow measurament from EKF
        zranger : ZRanger = None
        # disable flow contribution
        if DeckType.bcFlow2 in ecf.decks:
            ecf.decks[DeckType.bcFlow2].
                 contribute_to_state_estimate = True 
        # get the ZRanger of the FlowDeck
                 zranger = ecf.decks[DeckType.bcFlow2].zranger 
        if DeckType.bcZRanger2 in ecf.decks:
            zranger = ecf.decks[DeckType.bcZRanger2]
            # disable zrange contribution
            zranger.contribute_to_state_estimate = True 
\end{lstlisting}

\section{Code Example}
\label{app:codeTeam5}

An excerpt of code produced by teams \emph{Flying Ferrets}. We use \code{\#...} to omit lines of code that are not relevant for the discussion. Because of formatting, we could not keep the regular Python indentation. The whole program is $\approx$ 380 lines of code.

\begin{lstlisting}
import logging
import sys
import time
# ...
import cflib.crtp  # to scan for Crazyflies instances
# to easily connect/send/receive data from a Crazyflie
from cflib.crazyflie import Crazyflie  
# wrapper around the normal Crazyflie class
from cflib.crazyflie.syncCrazyflie import SyncCrazyflie  
# to help connecting to a Crazyflie with a URI
from cflib.utils import uri_helper  
from Utility.log import Log  # to log position

# to help moving the drone
from cflib.positioning.motion_commander import MotionCommander  
# to use the Multiranger deck
from cflib.utils.multiranger import Multiranger  

#...

ADJUST_VELOCITY = 0.2 # [m/s]
threshold = 0.1 # [m]
DEFAULT_HEIGHT = 0.2
prev = 0 # 
class ActionLimit():
    #measure unit millimeters
    MIN = 0.150
    MAX = 0.400
    SAFE = 0.100

class VelocityLimit():
    #measure unit meters/second
    MIN = 0
    MAX = 0.8

def is_close(range, distance):
    if range is None:
        return False
    else:
        return range < distance

def is_far(range, distance):
    if range is None:
        return False
    else:
        return range > distance
    
def compute_velocity(value, limits=ActionLimit) -> float:
        #fixing values in the range (0, ACTION_LIMIT)
        value = ActionLimit.MIN if value < ActionLimit.MIN
                                else value
        value = ActionLimit.MAX if value > ActionLimit.MAX
                                else value
        # NewValue = (((OldValue - OldMin) * (NewMax - NewMin))
                                #/ (OldMax - OldMin)) + NewMin where:
            # OldValue = range_in_mm
            # NewValue = velocity_in_ms
            # OldMin = ActionLimit.MAX (range)
            # OldMax = ActionLimit.MIN (range)
            # NewMin = VelocityLimit.MIN (velocity)
            # NewMax = VelocityLimit.MAX (velocity)
        # NOTICE: we inverted the ActionLimits to get
        # the inverted range conversion
        return (((value - ActionLimit.MAX) *
                         (VelocityLimit.MAX - VelocityLimit.MIN)) /
                         (ActionLimit.MIN - ActionLimit.MAX)) +
                         VelocityLimit.MIN

# this function will compute the Velocity in the x-axis direction
def get_vx(front, back, limits=ActionLimit)-> float:
    vx = 0
    if(ActionLimit.MIN <= back <= ActionLimit.MAX):
        #back gives a push in the positive x direction
        vx += compute_velocity(back)
    if(ActionLimit.MIN <= front <= ActionLimit.MAX):
        #back gives a push in the negative x direction
        vx -= compute_velocity(front)
    return vx
# this function will compute the Velocity in the y-axis direction
def get_vy(right, left, limits=ActionLimit)-> float:
    vy = 0
    if(ActionLimit.MIN <= right <= ActionLimit.MAX):
        #back gives a push in the positive y direction
        vy += compute_velocity(right)
    if(ActionLimit.MIN <= left <= ActionLimit.MAX):
        #back gives a push in the negative y direction
        vy -= compute_velocity(left)
    return vy

def main():
    # Initialize the low-level drivers
    cflib.crtp.init_drivers()

    #... 

    cf = Crazyflie(rw_cache='./cache')
    with SyncCrazyflie(URI, cf=cf) as scf:
        print('SyncCrazyflie open')
        # to start logging position (DO NOT REMOVE)
        with Log(scf) as log:  

            # ...
            with MotionCommander(scf,
                        default_height=DEFAULT_HEIGHT)
                as motion_commander:
                print('MotionCommander open')
                with Multiranger(scf) as multiranger:
                    print('Multiranger open')
                    keep_flying = True
                    is_started = False
                    velocity_x = 0.0
                    velocity_y = 0.0
                    
                    lastt = time.time()
                    time.sleep(1)
                    while keep_flying:
                        #vz = compute_velocity
                            #(multiranger.down, 10000.0)
                        #print(multiranger.front, multiranger.back,
                            #multiranger.right, multiranger.left)

                        front = multiranger.front
                        back = multiranger.back
                        left = multiranger.left
                        right = multiranger.right
                        if front is None:
                            front = 1000.0
                        if back is None:
                            back = 1000.0
                        if left is None:
                            left = 1000.0
                        if right is None:
                            right = 1000.0

                        if is_close(multiranger.down,
                                    DEFAULT_HEIGHT - threshold):
                            vz = ADJUST_VELOCITY*2
                            #motion_commander.
                                #start_linear_motion(0, 0,
                                #ADJUST_VELOCITY)
                        elif is_far(multiranger.down,
                                    DEFAULT_HEIGHT + threshold):
                            vz = -ADJUST_VELOCITY
                            #motion_commander.
                                #start_linear_motion(0,0,
                                #-ADJUST_VELOCITY)
                        else:
                            vz = 0
                            #motion_commander.
                            #start_linear_motion(0, 0, 0)

                        if multiranger is None:
                            print("Multiranger is None!")
                            motion_commander.land()
                            time.sleep(0.5)
                            break
                        
                        if time.time()-lastt > 0.1:
                            print(f"{front} {back} {right} {left}")

                        vx = get_vx(front, back)
                        vy = get_vy(right, left)
                        if time.time()-lastt > 0.1:
                          print(f"{mr.down} {vx} {vy} {vz}")
                          lastt = time.time()
                        motion_commander.
                             start_linear_motion(vx, vy, vz)
                        time.sleep(0.01)

if __name__ == "__main__":
    main()
\end{lstlisting}

\end{document}